\documentclass[10pt,a4paper]{npqcd}
\begin{document}
\renewcommand\nextpg{\pageref{pgs1}}\renewcommand\titleA{Bimodality as a signal of the nuclear liquid-gas phase transition
}
\renewcommand\authorA{
 V. V. Sagun${}^{\mathrm{a}}$,
 A. I. Ivanytskyi${}^{\mathrm{b}}$,
 D. R. Oliinychenko${}^{\mathrm{c}}$,
 K. A. Bugaev${}^{\mathrm{d}}$
}\renewcommand\email{
 e-mail:\space \eml{a}{v\underline{ }sagun@ukr.net}, \eml{b}{a\underline{ }iv\underline{ }@ukr.net},
  \eml{c}{dimafopf@gmail.com}, \eml{d}{bugaev@th.physik.uni-frankfurt.de}\\[1mm]
}\renewcommand\titleH{
Bimodality as a signal of the nuclear liquid-gas phase transition
}\renewcommand\authorH{
 Sagun ~V.~V., Ivanytskyi ~A.~I., Oliinychenko ~D.~R., Bugaev ~K.~A.
}\renewcommand\titleC{\titleA}\renewcommand\authorC{\authorH}\renewcommand\institution{
 Bogolyubov Institute for Theoretical Physics, Kiev, Ukraine
 }
\renewcommand\abstractE{
Here we present an explicit  counterexample to  a  bimodality concept  as the unique signal of  first order phase transition.
Using an exact  solution of  the simplified version of the statistical multifragmentation model we demonstrate that  the bimodal distributions can naturally appear  in infinite  system without a phase transition in the regions of the negative values of the surface tension coefficient.
Also we propose a new parameterization for the compressible nuclear liquid which is consistent with the L. van Hove axioms  of statistical mechanics. As a result the proposed model  does not lead to the irregular  behaviour of the isotherms in the mixed phase region which  is typical for mean-field models.
Peculiarly, the suggested approach to account for the nuclear liquid compressibility automatically leads to an appearance of an additional state that in many respects resembles the physical antinuclear matter.
}

\begin{article}
\section{Introduction}

At the present time the  bimodality is often considered as a signal of the first order PT in finite systems. The authors  of such schemes  \cite{Sagun:Chomaz03,Sagun:Gulm07} identify each  local maximum of the bimodal distribution with a pure phase. Such an idea goes back to T. Hill   book  \cite{Sagun:THill1}. Hill   justified  this assumption on bimodality by stating that due to the fact that an interface between two pure phases 'costs' some additional energy, the probability of their coexisting in a finite system is less than for each of pure phases \cite{Sagun:THill1}. 
It was found \cite{Sagun:Bugaev07}, however, that such an assumption can be  valid for infinite systems only.
In order to demonstrate that Hill  assumption can be incorrect even in thermodynamic limit, 
 here we  present a clear  counterexample by considering an exact analytical solution of the constrained statistical multifragmentation model (CSMM) in  thermodynamic limit which leads to the bimodal fragment size distributions inside of  the cross-over  region. For this purpose we consider a more realistic equation of state for the liquid phase which, in contrast to the original SMM formulation \cite{Sagun:Bugaev00}, is a compressible one \cite{Sagun:Bugaev13}. The second important element of the present model  is a more realistic  parameterization of  the temperature dependent  surface tension  based on an exact analytical solution of  the  partition function of surface deformations \cite{Sagun:HDM}.

\subsection{CSMM with compressible nuclear liquid in thermodynamic limit}

The general solution of the CSMM  partition function formulated in the grand canonical variables of volume $V$, temperature $T$ and baryonic chemical potential $\mu$ is given by \cite{Sagun:CSMM05}
\begin{equation}\label{Sagun:I}
{\cal Z}(V,T,\mu)~ = \sum_{\{\lambda _n\}}
e^{\textstyle  \lambda _n\, V }
{\textstyle \left[1 - \frac{\partial {\cal F}(V,\lambda _n)}{\partial \lambda _n} \right]^{-1} } \,,
\end{equation}
where  the set of  $\lambda_n$ $(n=0,1,2, 3,..)$ are all the complex roots of  the equation 
\begin{equation}\label{Sagun:II}
\lambda _n~ = ~{\cal F}(V,\lambda _n)\,,
\end{equation}
ordered as   $Re(\lambda_n) > Re(\lambda_{n+1})$ and $Im (\lambda_0) = 0$. The function ${\cal F}(V,\lambda)$ is defined as 
\begin{eqnarray}\label{Sagun:III}
&&\hspace*{-0.6cm}{\cal F}(V,\lambda)=\left(\frac{m T}{2 \pi}\right)^{\frac{3}{2}}z_1\exp \left\{\frac{\mu-\lambda T b}{T}\right\}+\hspace*{-0.1cm} \sum_{k=2}^{K(V)}\phi_k (T) \exp \left\{\frac{( p_l(T,\mu)- \lambda T)b k }{T} \right\}\,.~~~~~
\end{eqnarray}
Here $m \simeq 940$ MeV is a nucleon mass, $z_1 = 4$ is an internal partition (the degeneracy factor) of nucleons, $b = 1/ \rho_0 $ is the eigen volume of one nucleon in a vacuum ($\rho_0\simeq 0.17$ fm$^3$ is the normal nuclear density at $T=0$ and zero pressure). The reduced distribution function of the $k$-nucleon fragment in (\ref{Sagun:III}) is defined as 
\begin{equation}\label{Sagun:IV}
 \phi_{k>1}(T)\equiv\left(\frac{m T }{2 \pi}\right)^{\frac{3}{2} } k^{-\tau}\,\exp \left[ -\frac{\sigma (T)~ k^{\varsigma}}{T}\right]\,,
\end{equation} 
where $\tau \simeq 1.825$ is the Fisher topological exponent and $\sigma (T)$ is the $T$-dependent surface tension coefficient.  Usually, the constant, parameterizing the dimension of surface in terms of the volume is  $\varsigma = \frac{2}{3}$.
\begin{figure}[hbt]
\centering
 \includegraphics[width=.575\textwidth]{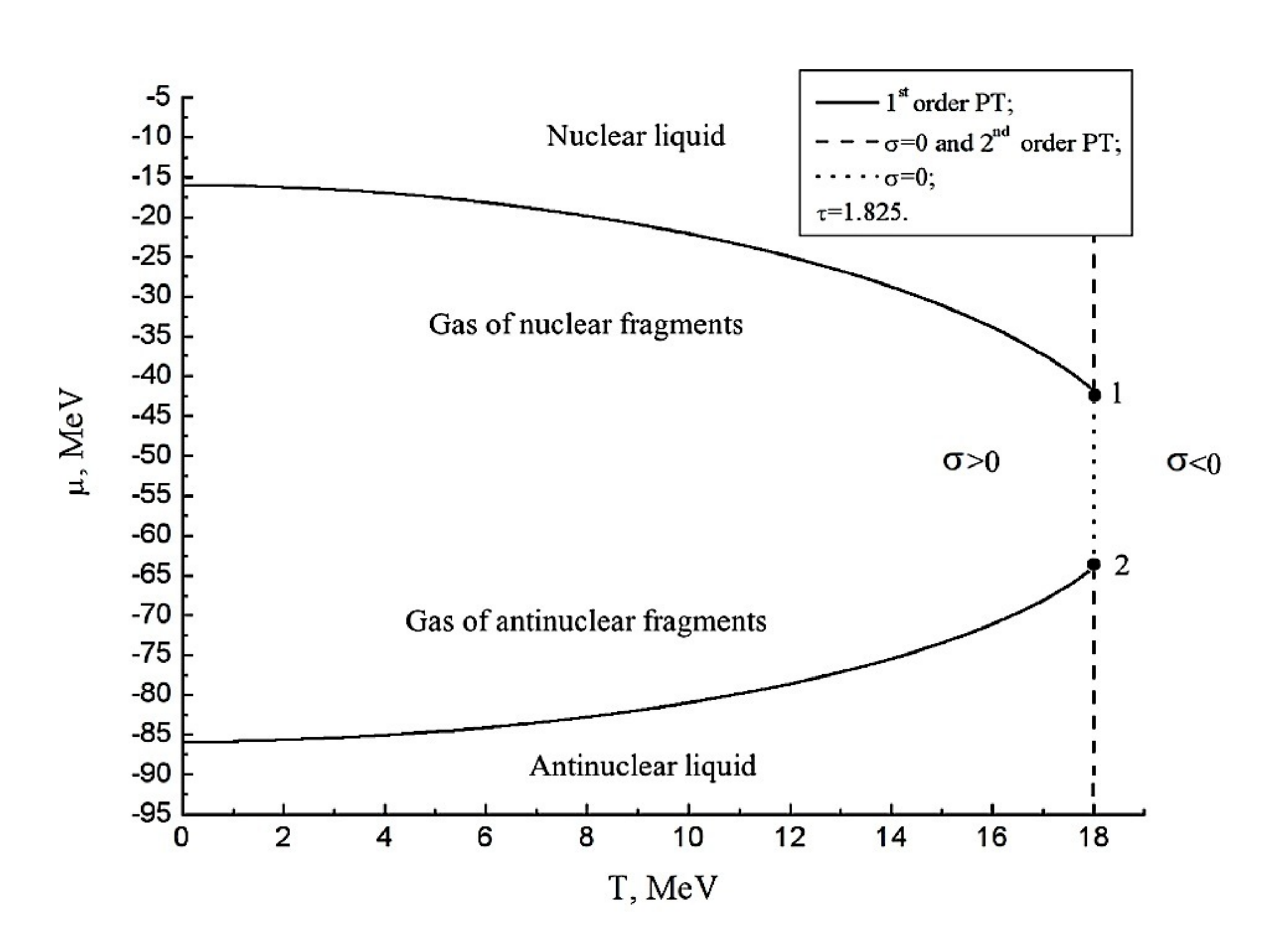}
 \includegraphics[width=.575\textwidth]{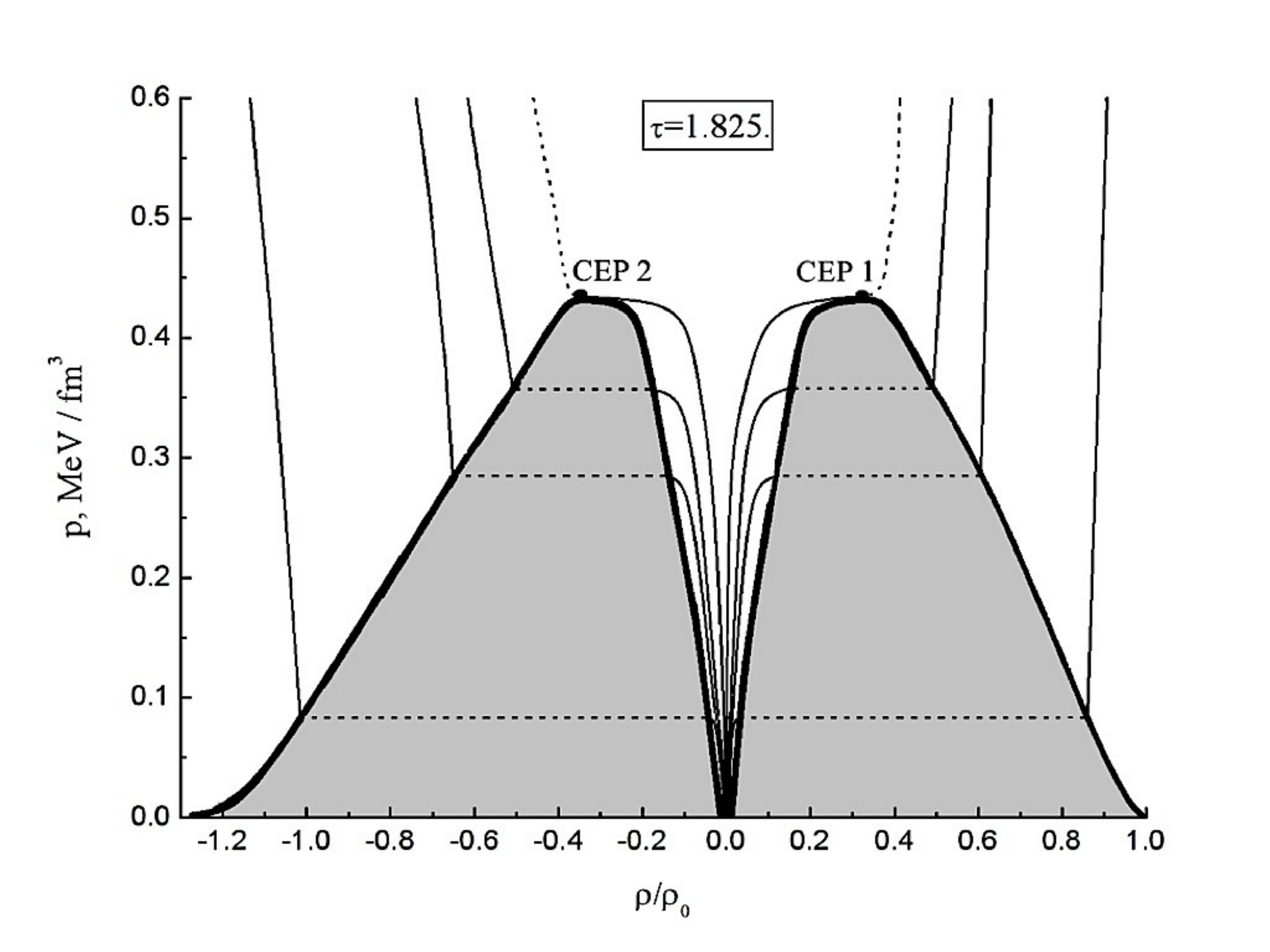}
  \caption{{\bf Upper panel:} The phase diagrams in $T-\mu$ plane. Along the solid curves there are first order PTs.The vertical dashed lines show the second order PT and the black circles correspond to the tricritical endpoints marked by the digits 1 (nuclear matter) and 2 (antinuclear matter). A cross-over occurs along the  dotted vertical line of the vanishing surface tension coefficient.
  {\bf Lower panel:} The phase diagrams in $\rho-p$ plane. The grey  areas show the mixed phases of the first  order PTs. The isotherms are shown for $T=11, 16, 17, 18$ MeV  from bottom to top. Negative density values correspond to an   `antimatter'. 
}
  \label{Sagun:fig:1}
\end{figure}

In (\ref{Sagun:III}) the exponentials  $\exp( - \lambda b k)$ ($k=1,2,3,...$) appear due to the hard-core  repulsion between the nuclear fragments \cite{Sagun:CSMM05}, while $p_l(T,\mu)$ is the pressure of the liquid phase. 
Here we consider the thermodynamic limit only, i.e. for $V \rightarrow \infty$  we have  $K(V) \rightarrow \infty$. Then the treatment of the model is essentially simplified, since Eq. (\ref{Sagun:II}) can have only two kinds of solutions \cite{Sagun:CSMM05}, either the gaseous pole $p_g (T, \mu) = T \lambda_0 (T, \mu)$ for  ${\cal F}(V,\lambda_0 - 0) < \infty$ or the liquid essential singularity 
$p_l (T, \mu) = T \lambda_0 (T, \mu)$ for  ${\cal F}(V,\lambda_0 - 0) \rightarrow  \infty$. The mathematical reason why  only the rightmost solution   $\lambda_0 (T, \mu) = \max \{Re(\lambda_n)\} $  of   Eq. (\ref{Sagun:II}) 
 defines the system pressure is evident from Eq. (\ref{Sagun:I}): in the limit $V \rightarrow \infty$ all  the solutions  of (\ref{Sagun:II}) other than the rightmost one are exponentially suppressed.  

In the thermodynamic limit the model has a PT, when there occurs a change of the  rightmost solution type, i.e. when the gaseous pole is changed by  the liquid essential singularity or vice versa. The PT line $\mu = \mu_c (T)$ is a solution of  the equation of  `colliding singularities' $p_g (T, \mu) = p_l (T, \mu) $, which is just the Gibbs criterion of  phase equilibrium. The properties of a PT are defined only by the liquid phase pressure  $p_l (T, \mu)$ and   by the temperature dependence of  surface tension $\sigma(T)$.

In order to consider the compressible  nuclear liquid in \cite{Sagun:Bugaev13} we suggested  the following  parameterization of its pressure 
\begin{eqnarray}
\label{Sagun:V}
p_l=\frac{ W(T) +  \mu + a_2 ( \mu -\mu_0)^{2} + a_4 ( \mu -\mu_0)^{4}}{b} \,.
\end{eqnarray}
Here $ W(T) = W_0 + \frac{T^2}{W_0}$ denotes  the usual  temperature dependent  binding energy per nucleon with $W_0 =  16$ MeV \cite{Sagun:Bugaev00} and  the constants  $\mu_0 = - W_0$, $a_2 \simeq 1.233 \cdot 10^{-2}$ MeV$^{-1}$ and $a_4 \simeq 4.099 \cdot 10^{-7}$ MeV$^{-3}$.  These constants  are fixed in order   to reproduce  the properties  of normal nuclear matter, i.e. at vanishing temperature  $T=0$ and normal nuclear density $\rho = \rho_0$ the liquid pressure must be zero. 

In addition to the new parameterization of the free energy of the $k$-nucleon fragment (\ref{Sagun:III}) we  consider a more general parameterization of the surface tension coefficient 
\begin{equation}\label{Sagun:IX}
 \sigma (T) =  \sigma_0 \left| \frac{T_{cep} - T }{T_{cep}} \right|^\zeta  {\rm sign} ( T_{cep} - T) ~,
\end{equation}
with  $\zeta = const \ge 1$, $T_{cep} =18$ MeV and $\sigma_0 = 18$ MeV the SMM. In contrast to the Fisher droplet model \cite{Sagun:Fisher67} and the usual SMM, the CSMM surface tension (\ref{Sagun:IX}) is negative above the critical temperature $T_{cep}$. An extended discussion on the validity of such a parameterization can be found in \cite{Sagun:Bugaev13}. The resulting  phase diagrams of the present model in different variables are shown in Fig.~\ref{Sagun:fig:1}.
\begin{figure}[h]
\hspace*{-0.44cm}\begin{minipage}{.59\textwidth}
\includegraphics[width=.88\textwidth]{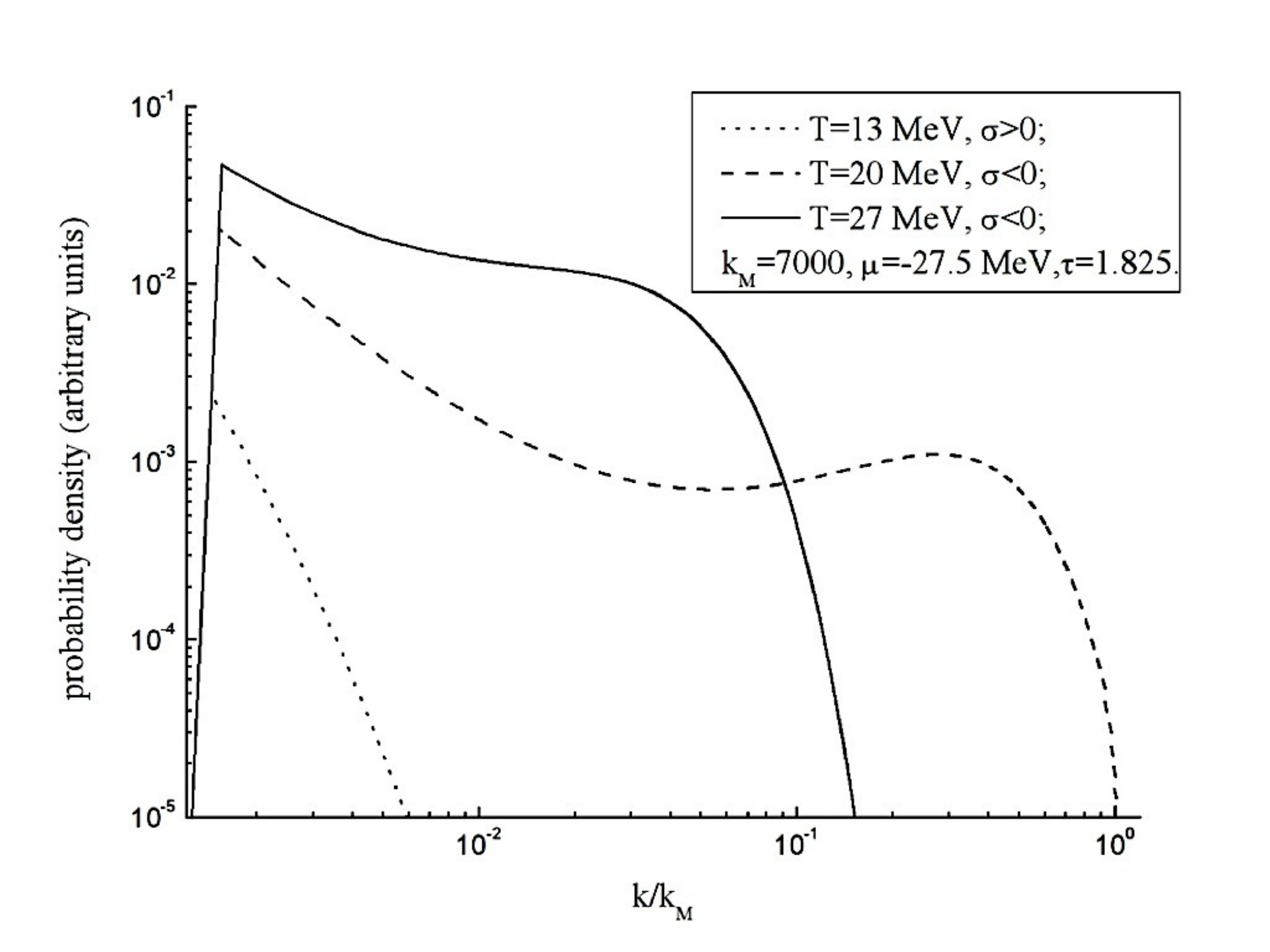}
\end{minipage}
\rule{-.11\textwidth}{0pt}
\begin{minipage}{.6\textwidth}
\includegraphics[width=.99\textwidth]{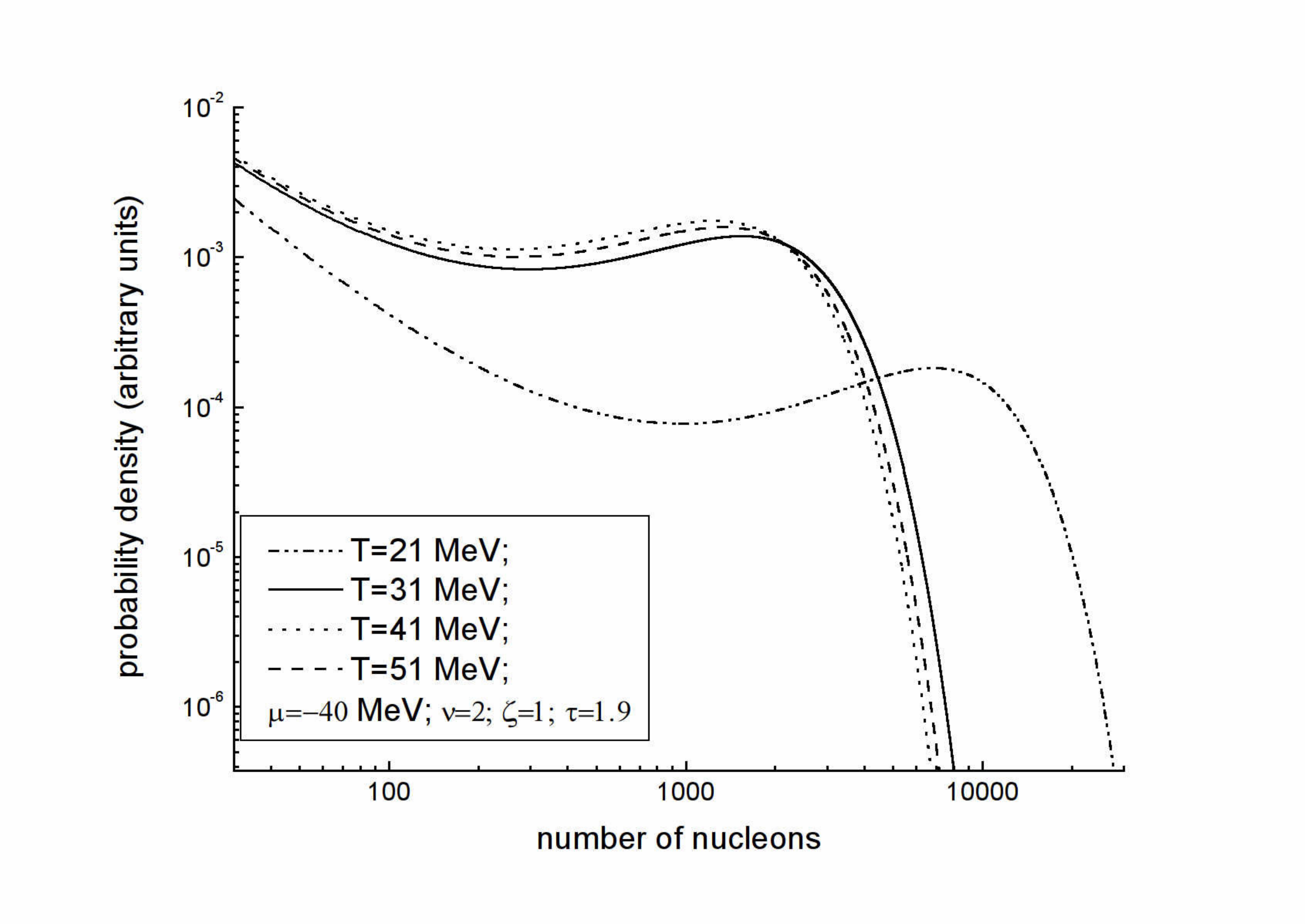}
\end{minipage}
 \caption{{\bf Left panel:}  Fragment size distribution of the model  is shown for a fixed baryonic chemical potential $\mu = -27.5$ MeV and three values of the  temperature $T$.  {\bf Right panel:}
 Same as  in the left panel, but for  a fixed baryonic chemical potential
and different temperatures located at the region of   negative values of the surface
tension coefficient and  $\nu$ = 2. 
}
 \label{Sagun:fig:2}
\end{figure}

In order to elucidate the role of the negative surface tension coefficient
  we  study the fragment size distribution in two regions  of the phase diagram.  
To demonstrate the pitfalls of the bimodal concept of Refs. \cite{Sagun:Chomaz03, Sagun:Gulm07,Sagun:THill1} we compare  the gaseous phase fragment  size distribution with that one in the supercritical temperature region, where there is  no  PT  by construction.  As one can see from Fig. \ref{Sagun:fig:2}  in the gaseous phase, even at the boundary with the mixed phase,  the size distribution is a monotonically   decreasing function of  the number of nucleons in a fragment $k$. However, for the supercritical temperatures one finds the typical bimodal fragment distribution for a variety of temperatures and chemical potentials as one can see from Fig. \ref{Sagun:fig:2}.

A sharp peak at low $k$ values reflects  a fast increase of the  probability density of dimers compared to the monomers (nucleons),  since the intermediate  fragment sizes do not have the binding free energy and the surface free energy and, hence, the monomers are significantly suppressed in this region of thermodynamic parameters. On the other hand it is clear that the tail of fragment distributions in Fig. \ref{Sagun:fig:2} decreases due to the dominance of the bulk free energy and, hence, the whole structure at intermediate fragment sizes is due a competition between the surface free energy and two other contributions into the fragment free energy, i.e. the bulk one and the Fisher one. 
It was also found that with temperature increasing the minimum and maximum of the
distribution function grow wider and shallower and they shift towards the smaller number
of nucleons in a fragment (see the right  panel of Fig. \ref{Sagun:fig:2}).
 
\subsection{Conclusions}

In the present work we showed that the bimodal distributions can naturally appear in an infinite system without a PT. 
Our analysis of the fragment size distributions in the region of negative surface tension coefficient shows that these distributions have a saddle-like shape. Such a behavior  closely resembles  the fragment size distribution  observed in dynamical simulations of nuclear multifragmentation  \cite{Sagun:Campi03}.
The compressible nuclear liquid pressure parametrization which generates the tricritical endpoint at the one third of the normal nuclear density is worked out.

\end{article}
\label{pgs1}


\begin{thebibliography}{11}\bibitemsep
 \bibitem{Sagun:Chomaz03} Ph. Chomaz and  F. Gulminelli,
 Preprint GANIL-02-19, (2002).

\bibitem{Sagun:Gulm07} F. Gulminelli,
 Nucl. Phys. A {\bf 791},  165,  (2007).
 
\bibitem{Sagun:THill1} T. L. Hill,  {\it Thermodynamics of small  systems} Dover, New York, 1994.

\bibitem{Sagun:Bugaev07}
%
K. A. Bugaev, 
Phys. Part. Nucl. {\bf 38}   447, (2007);
 arXiv:nucl-th/0511031. 

\bibitem{Sagun:Bugaev00}
%
K. A. Bugaev,
M. I. Gorenstein, I. N. Mishustin and W. Greiner,
Phys. Rev. C {\bf 62},  044320, 2000;
arXiv:nucl-th/0007062;
%
Phys. Lett. B {\bf  498},  144, 2001;
arXiv:nucl-th/0103075.

\bibitem{Sagun:Bugaev13}
K. A. Bugaev, A. I. Ivanytskyi, V. V. Sagun and D. R. Oliinychenko, 
Phys.  Part. Nucl. Lett. {\bf 10}, 832 (2013);
arXiv:1306.2481  [nucl-th].
  
\bibitem{Sagun:HDM}
%
K. A. Bugaev, L. Phair and J. B. Elliott,
Phys. Rev. E {\bf  72}, 047106  (2005); 
 arXiv:nucl-th/0406034. 
 

\bibitem{Sagun:CSMM05}
%
K. A. Bugaev,
Acta. Phys. Polon. B {\bf  36}, 3083, (2005) and reference therein.

\bibitem{Sagun:Fisher67}
M. E. Fisher, Physics {\bf 3},  255, (1967).

\bibitem{Sagun:Campi03}
%
X. Campi,  H. Krivine  E. Plagnol and N. Sator, 
Phys. Rev. C {\bf 67}, 044610, (2003).

\end{thebibliography}
\end{document}